\documentstyle[epsf]{article}

\newcommand{\tr}{ \hbox{tr} }
\newcommand{\be}{\begin{eqnarray}}
\newcommand{\ee}{\end{eqnarray}}

\begin{document}

\title{
 Nuclear Clusters as a Probe for \\
Expansion Flow  \\
 in 
 Heavy Ion Reactions at 10-15AGeV
       }
\author{
R. Mattiello $^{1,2}$\\ 
  H. Sorge $^2$\\
 H. St\"ocker $^3$, 
  and W. Greiner $^3$ 
\\
$^{1}$
Physics Department, Brookhaven National Laboratory\\
Upton, New York 11973
\\
$^{2}$
Physics Department, S.U.N.Y. at Stony Brook\\
 Stony Brook, NY 11974
\\
$^{3}$ 
Institut f\"ur Theoretische Physik, 
 J.W. Goethe Universit\"at\\
D-60054 Frankfurt am Main, Germany}  
%
%
\maketitle  
  
\begin{abstract}  
A phase space coalescence description 
based on the Wigner-\-function method
for cluster formation in relativistic 
nucleus-nucleus collisions is presented.
The  momentum distributions of nuclear clusters
d,t and He are predicted  for central Au(11.6AGeV)Au
and Si(14.6AGeV)Si reactions in the framework
of the RQMD transport approach.
Transverse expansion  
leads to a strong shoulder-arm shape and different
inverse slope parameters in the transverse spectra of
nuclear clusters  deviating markedly from thermal distributions. 
A clear ``bounce-off'' event shape is seen: 
the averaged transverse flow velocities 
in the reaction plane are for clusters 
larger than for protons.  
The cluster yields --particularly at low $p_t$ at midrapidities--
and the in-plane (anti)flow of clusters
and pions change 
if suitably strong baryon  potential interactions are included.
This allows to study the transient pressure at high density 
via the event shape analysis of nucleons, nucleon clusters
and  other hadrons.
\end{abstract}   
%

\section{Introduction}
One of the challenges of modern heavy ion physics is the
extraction of the equation of state and transport properties of
 extremely dense and excited
nuclear matter. In particular,
the study of matter at
high net baryon density has received much attention recently.
QCD - as the accepted theory of
strong interaction - contains chiral symmetry (in the limit
of massless quarks) which is spontaneously broken in its ground state, 
the QCD vacuum (see e.g. lattice calculations \cite{19}).
A rapid restoration of this symmetry with increasing
baryon density is predicted by all approaches which embody this
fundamental aspect of QCD \cite{20,7}. 
Nucleus-nucleus collisions in the bombarding energy region of
baryon stopping may therefore be favourable to study such medium effects
as compared to ultrahigh energies, for which the two colliding nuclei
may become transparent. 
Beam energies between
10 to 15  AGeV -- as studied experimentally at the BNL-AGS
\cite{2}-\cite{trenerg} --
seem to be well suited to stop two heavy ingoing nuclei
and to create the desired high baryon densities.
This has been shown by transport calculations based on 
hadronic excitations and rescattering like the RQMD approach 
(strings, resonances) \cite{9,8}, the ARC- \cite{22} or the ART-model 
\cite{ART} (resonances).
The observation of stopping in the AGS experiments
has been  unclear for quite some time.
However, all experimental groups 
now confirm \cite{2}-\cite{E814}
the predicted large baryon stopping in central
collisions \cite{9,Keitz}. 

An observable consequence of the formation of dense nuclear matter
-- far beyond the groundstate -- is the emergence of collective
flow driven by  compression-induced pressure
 \cite{21}-\cite{casflow}.  
Mean fields \cite{21} may give 
important contributions to this pressure and could 
therefore be accessible to experimental observation, 
just as in the 1GeV region \cite{experiment}.
The bounce-off for protons has been observed at
10 GeV/n \cite{BMbodrum}
as well as azimuthally asymmetric particle correlations in the 
projectile hemisphere \cite{asym}.
These  experimental discoveries encourage us to 
investigate the formation of nuclear clusters
 -- as compared to light hadrons -- 
for which flow can even dominate the momentum spectra \cite{ogloblin}.

We follow our earlier work on deuterons \cite{di18}
and extend the phase space coalescence picture to
light clusters with $A\le 4$.
The basic ingredients 
of the cluster coalescence, i.e. 
the source function provided by the RQMD and
the parametrization of cluster wave functions
are described in Sect. 2 and 3. The parametrization of baryonic
mean fields is described in Sect. 4. In Sect. 5
particular features in the momentum distributions of nuclear clusters
are discussed as signatures for flow 
and event shape correlations at the hadronic freeze-out.
In order to demonstrate sensitivities to baryonic mean fields
final observables --rapidity distributions, $m_t$-spectra,
directed flow $p_x(y)$-- are compared for  
two extreme scenarios: one with a density-dependent
quasi-potential between baryons,
and the other without (cascade). 

It should be mentioned that
results for cluster yields have been calculated based on
the thermal model and on the single-nucleon momentum
distributions 
\cite{di117,di118,di119,coal1,di134a,di134b,di134c,di75,di131,di132,di133}.
These results predict that the spectra of clusters and
nucleons have essentially the same shape. Here, we
are going to demonstrate that flow invalidates the
basic assumptions underlying these simple models.
In turn, we can use the amount of `scaling violation'
of cluster spectra as compared to proton spectra to assess the strength
of collective flow in nucleus-nucleus reactions.
  
\section{Coalescence of Clusters in Phase Space}
 We combine a dynamical description
of the first violent stage in nucleus-nucleus reactions
 from the RQMD model with a cluster
formation model which is based on the single-particle
phase space distributions at freeze-out.
RQMD is a semi-classical transport theoretical approach and
does not take into account  formation of nuclear bound states
(e.g. deuterons) dynamically.
However, the small binding energies 
and the associated
quantum mechanical formation time from the uncertainty principle
suggest that
nuclear clusters are  produced mainly much later
after the violent interactions
have ceased (freeze-out), i.e.
 cluster formation rates can be calculated
from the nucleon distributions at freeze-out.
In order to calculate light nuclear cluster distributions
(for A$\le$4) we 
use the Wigner-function technique
in phase space.
This phase space coalescence approach 
 was already  applied to deuteron production at
bombarding energies around 1AGeV
 \cite{di128,di129},   10-15AGeV \cite{di18,jamie,bleicher}
 and 160-200AGeV \cite{clscaling}.

The validity of the combined RQMD+coalescence
approach for the cluster formation clearly
depends on that 
the transport model 
describes the
dynamical evolution
up to the freeze-out stage reasonably well. 
The relativistic quantum molecular dynamics approach
(RQMD 1.07) \cite{9} employed for the calculations presented here
combines the classical propagation
of particles with  excitation of hadrons into 
resonances and strings. 
Secondaries (emerging from the decaying resonances and
 strings) undergo subsequent interactions, both
with each other and with the ingoing baryons.
Note that the RQMD-results compare well with experimental single 
particle and 2-body correlation data \cite{E802,E814,8,correl}.

In the following,
 we use the non-relativistic Wigner-function formalism
 which may be  justified for  nuclear clusters
 in view of the small binding energies.
The  formation rates are calculated at equal time in the 
common rest frame of the corresponding cluster nucleons 
immediately after their freeze-out.
Having this in mind, we suppress the explicit time dependence 
and reference to the chosen Lorentz frame in 
all following expressions. 
Note, however, 
that all results presented in this work include an implicit
integration over all freeze-out times
and Lorentz transformations back into the original
observer system.

The Wigner-function for a single particle
\be 
\int d^3y <\vec x+{\vec y/2}|\Psi>
               <\Psi|\vec x-{\vec y/2}> e^{-{i\over\hbar}\vec p\cdot\vec y}
\ee
is the closest analogue to a 
 probabilistic distribution in phase space which one
 can get from quantum mechanics. Therefore, its
identification with the phase space distribution 
$f_N\approx\rho^W$ has been frequently
 employed in semi-classical calculations.
 Neglecting the hopefully small effects from binding energies,
 the formation probability for a cluster can be expressed
as an overlap integral between the
Wigner-function which corresponds to the cluster wave function
  and the N-body nucleon phase space distribution at freeze-out
\cite{di127,di128,di129,di18,jamie,clscaling}.
The N-body phase space distribution has to be constructed from
the single particle ``source function'' which is defined by the  
``freeze-out'' positions
$x_i^\mu$ and momenta $p_i^\mu$ of nucleons
after their last scattering or decay.

The Wigner-density of an M-nucleon state has the form
\be
\label{eqdens}
[\hat \rho_M]^W &=& \sum_{[TT_3]\times[SS_3]} {(|SS_3><SS_3|)} 
             \times {(|TT_3<TT_3|)}\nonumber\\
             && \times W_{[TT_3]\times[SS_3]}(x_1,p_1;...,x_M,p_M)
\ee
with the normalization $\tr[\hat\rho_M]=1$. The product
$[TT_3]\times[SS_3]$ denotes the set of all $2^{2M}$
possible internal couplings with proper total isospin $T,T_3$ 
and spin $S,S_3$. Note that the phase space part 
$W_{[TT_3]\times[SS_3]}$ has to provide the correct symmetry 
concerning particle exchange to ensure that all states 
are totally antisymmetric. 
In the  semiclassical approximation it is assumed that 
the Wigner-function  does not contain $dynamical$ correlations 
with respect to spin and isospin. 
We therefore employ the statistical assumption 
and assign all many-nucleon states           
which are allowed from the Pauli principle  the same 
weight for a given  position and momentum distribution:


We consider only  spin averaged Wigner-densities
$W_{[TT_3]\times[SS_3]}\approx1/2^M \tilde W_{[T,T_3]}$.
Furthermore, the coupling of $M$ particles to a  given total isospin $T$ is
assumed to be equal
\be  
\label{eqmast}
W_{[TT_3]\times[SS_3]}\approx{1\over {M\choose Z}}{1\over2^M} \bar W_{T_3}
= g \; \bar W_{T_3}
\ee
In this approach $\bar W_{T_3}$ contains all spin states and  
all ${M\choose Z}$ states in different isospin-multiplets for a
M-particle combination $(M,Z)$ with given charge $Z=T_3+M/2$.

In order to approximate $\bar W_{T_3}$ we use the RQMD model 
that provides the phase space distribution of nucleons
with given isospin.  We identify $\bar W_{T_3}$ with the product of
single particle distributions
\be
\label{eqapp} 
\bar W_{T_3}\approx 
{1\over N^M}{M\choose Z} \left[\prod_{i=1}^Z (2\pi\hbar)^3 
                  f_p(\vec x_i,\vec p_i)\right]
                  \times \left[\prod_{i=Z+1}^M(2\pi\hbar)^3 
                  f_n(\vec x_i,\vec p_i)\right]
\ee
where $N=N_p+N_n$ is the total number of nucleons and 
\be 
     N_p:=\int {d^3xd^3p} \; f_{p}(\vec x,\vec p) \quad,\quad
     N_n:=\int {d^3xd^3p} \; f_{n}(\vec x,\vec p)
\ee
Eq. \ref{eqapp} can be interpreted as a statistically uncorrelated 
emission. It defines the probability-density to find
a given nucleon combination (M,Z) in certain phase space regions. 
Inserted in Eq. \ref{eqmast} it
fulfills by construction the trace normalization $\tr{[\hat\rho_M]}=1$.

The cluster wave function  
which
is assumed to be non-relativistically here
factorizes into a collective and
a relative part 
\be
|\Psi_C(\vec P)>
={1\over(2\pi\hbar)^{3/2}} e^{{i\over \hbar}\vec P\cdot \vec X} 
                      \phi(\vec t_1,...,\vec t_{M-1})
                                  |SS_3> |TT_3>       
\ee 
where $\vec X=(\vec x_1+...+\vec x_M)/M$
and $\vec P=\vec p_1+...+\vec p_M$. 
The $\vec t_i(\vec x_1,...,\vec x_M)$ (i=1,...,M-1)
are the $M-1$ relative coordinates of the 
relative cluster wave function $\phi$. $S,S_3,T,T_3$ are the spin and
isospin quantum numbers of the cluster state.
The Wigner-density of the wave function in relative coordinates
is defined by the Wigner-transformed projection operator  
\be
 \hat\rho_C^W(\vec t_1,\vec q_1;...;\vec t_{M-1},\vec q_{M-1}) := 
  \quad |TT_3>|SS_3>\rho_C^W<SS_3|<TT_3| 
\ee
with
\be
\rho_C^W&:=& \int     \phi(\vec t_1+\vec y_1/2
                ,...,\vec t_{M-1}+\vec y_{M-1}/2) 
                   \phi^*(\vec t_1-\vec y_1/2
                   ,...,\vec t_{M-1}-\vec y_{M-1}/2) \nonumber\\
 &&\times e^{-{i\over \hbar}\vec q_1\cdot\vec y_1} \; ... \; 
e^{-{i\over \hbar} \vec q_{M-1}\cdot\vec y_{M-1}} 
d^3y_1 \; ... \; d^3y_{M-1} \label{eqclust}
\ee
The formation of cluster states is finally determined by the trace 
over the source density $\hat\rho_M$ and the 
projector on the individual cluster wave function $|\Psi_C^M><\Psi_C^M|$:

\be
\label{eqint}
{\rm tr}\{\hat\rho_M|\Psi_C^M><\Psi_C^M|\}&=& 
\int [\hat\rho_M]^W(\vec x_1,\vec p_1;...;\vec x_M,\vec p_M) 
\hat\rho_C^W(\vec t_1,\vec q_1;...;\vec t_{M-1},\vec q_{M-1})\nonumber\\
&& \qquad\times\delta^3(\vec P-(\vec p_1+...+\vec p_M)) 
               {dx^3_1d^3p_1\over (2\pi\hbar)^3} ... 
               {d^3x_Mdp^3_M\over (2\pi\hbar)^3}
\ee
The absolute number of states is obtained by 
multiplying  Eq. \ref{eqint} with 
the total number of M-nucleon states ${N\choose M}$
and summation over all possible spin states $N_S$.
Inserting Eq. \ref{eqdens} and \ref{eqclust} 
the semi-classical coalescence formula reads finally
\be
{dN\over d^3P}&=&
g N_S \; {N\choose M} \; {M\choose Z} \; {1\over N^M} \; 
\int dx_1^3dp_1^3...dx_M^3dp_M^3 \delta^3(\vec P-(\vec p_1+...+\vec p_M))
\nonumber \\ 
&& \times\left[\prod_{i=1}^Z
 f_p(\vec x_i,\vec p_i)\right]
   \left[\prod_{i=Z+1}^M f_n(\vec x_i,\vec p_i)\right] 
   \rho_C^W(\vec t_1,\vec q_1;...;\vec t_{M-1},\vec q_{M-1})
\ee
For d, t, ${^3}$He 
and ${^4}$He states the momentum distributions are explicitly given by 
\be
{dN(d)\over d^3P}&=&
g({\rm d}) \; N_S({\rm d})\; {N\choose 2} {2\choose 1}\nonumber\\
&&\times{1\over N^2}
\int dx_1^3dp_1^3dx_2^3dp_2^3 
f_n(\vec x_1,\vec p_1)f_p(\vec x_2,\vec p_2)\nonumber\\
&& \times \rho_D^W \; 
                    \delta\left(\vec P-(\vec p_1+\vec p_2)\right)\\
{dN({\rm t})\over d^3P}&=&
g(t)\; N_S({\rm t})\; {N\choose 3} {3\choose 1}\; 
\nonumber\\
&& \times{1\over N^3} 
 \int dx_1^3dp_1^3...dx_3^3dp_3^3 
f_n(\vec x_1,\vec p_1)f_n(\vec x_2,\vec p_2)f_p(\vec x_3,\vec p_3)\nonumber\\
&& \times\rho_t^W \;
                    \delta\left(\vec P-(\vec p_1+\vec p_2+\vec p_3)\right)\\
{dN(^3{\rm He})\over d^3P}&=&
g(^3{\rm He})\;N_S(^3{\rm He})\; {N\choose 3}\;{3\choose2}\;
\nonumber\\
&&\times {1\over N^3} 
  \int dx_1^3dp_1...dx_3^3dp_3^3 
f_n(\vec x_1,\vec p_1)f_p(\vec x_2,\vec p_2)f_p(\vec x_3,\vec p_3)\nonumber\\
&& \times \rho_{^3{\rm He}}^W \;
                    \delta\left(\vec P-(\vec p_1+\vec p_2+\vec p_3)\right)\\
{dN(^4{\rm He})\over d^3P}&=&
g(^4{\rm He})\;N_S(^4{\rm He})\; {N\choose 4}{4\choose2}\;
\nonumber\\&&\times{1\over N^4}\int dx_1^3dp_1^3...dx_4^3dp_4^3 
f_n(\vec x_1,\vec p_1)f_n(\vec x_2,\vec p_2)
f_p(\vec x_3,\vec p_3)f_p(\vec x_4,\vec p_4)\nonumber\\
&&\times \rho_{^4{\rm He}}^W \;
            \delta\left(\vec P-(\vec p_1+\vec p_2+\vec p_3+\vec p_4)\right)
\ee
In a Monte Carlo formulation -- appropriate for the application
to microscopic transport calculations -- these formation rates can be 
expressed by the general coalescence formula for M-body cluster 
\be
dN_M &=& g N_S \left\langle \sum_{i_1,...,i_M\atop i_1<...<i_M} 
\rho_C^W(\vec t_{i_1},\vec q_{i_1};...;\vec t_{i_{M-1}},\vec q_{i_{M-1}})
)\right\rangle  \nonumber\\
&& \times  d^3t_{i_1}d^3q_{i_1}...d^3t_{i_{M-1}}d^3q_{i_{M-1}}
\quad. 
\label{gl374}
\ee
$<...>$ denotes event averaging. 
The  sum runs for each event over all M-nucleon combinations.
Note the necessary condition $i_1<...<i_M$ which prevents the
double counting of equal particle pairs.
The coordinates in position and momentum space  
are taken at equal time in the  M-nucleon rest frame
(i.e. $\vec P\equiv\vec0$) immediately after all cluster 
nucleons have frozen out.
The calculated numbers contain
higher mass fragments by construction. The number of A$>4$ clusters
is small, however, for rapidity values $|y-y_{mid}|<1$.
The factor $g$ contains spin and isospin projection as described above.
After the summation over all possible spin states the
statistical corections are $g({\rm d})N_S({\rm d})=3/8$, 
$g({\rm t})N_S({\rm t})=g(^3{\rm He})N_S(^3{\rm He})=1/12$ 
and $g(^4{\rm He})N_S(^4{\rm He})=1/96$.
Feed down effects from the production of excited t and He states
are expected to be small ($<15\%$, \cite{feedd}) and will be neglected in the 
present studies.

The statistical approximation employed here is expected to
break down in regions where the binding energies and the quantum 
dynamics play an essential role, e.g. in case of spectator matter 
fragmentation. Deviations from the statistical limit could give further 
insight into the fermionic (quantum) dynamics of the many-body system and 
final state effects like e.g. Coulomb distortion.

\section{The Parametrization of Cluster Wave Functions}
For the deuteron we assume a
Hulth\'en-wave function derived from
a Yukawa-type potential interaction \cite{f20,f22}
\be <\vec x_1,\vec x_2|D>
               &=&{1\over (2\pi\hbar)^{3/2}} 
             \exp\left({i\over\hbar}\vec P\cdot{\vec x_1+\vec x_2\over 2}
             \right)\nonumber 
               \times  {4ab(a-b)\over(a+b)^2}  \nonumber \\
               &&\times  {1\over |\vec x_1-\vec x_2|} 
            [\exp(-a|\vec x_1-\vec x_2|)-\exp(-b|\vec x_1-\vec x_2|)]. \ee
In order to get a simple expression for 
the Hulth\'en Wigner-density the
wave function is approximated
by a sum over 15 centrally symmetric Gaussian wave packets

\be \Psi(\vec r)=\sum_i a_i G(\vec r)=\sum_i a_i 
  \left({2c_i\over\pi}\right)^{3/4} \exp(-c_ir^2). \ee
The Wigner-density of this sum can be calculated analytically 
\cite{di149,di140}
\be \rho_D(\vec r,\vec q)
&=&8\sum_i a_i^2\exp(-2c_ir^2)\exp\left({q^2\over2c_i}\right) \nonumber \\
   && + 16 \sum_{i>j} a_i a_j
           \left({4c_ic_j\over(c_i+c_j)^2}\right)^{3/4} \nonumber \\
&&\times \exp\left({4c_ic_j\over c_i+c_j}r^2\right) 
\exp\left({-q^2\over c_i+c_j}\right)
\cos\left(2{c_i-c_j\over c_i+c_j} 
         r q \cos\theta_{rq}\right)
\ee 
with $q=|\vec p_1-\vec p_2|/2\quad,
\quad r=|\vec x_1-\vec x_2|$ as the relative position and
momentum coordinates.

For triton, $^3$He and $^4$He states we use 
3- and 4-particle harmonic oscillators with different
coupling strength. Such an approximation has been used already 
in momentum coalescence studies \cite{di118,di134a,di134b,di134c} 
and is well known
in nuclear structure physics (see e.g. \cite{di144}). 
We adopt such a form of the wave function because of the separability in
collective and relative motion even on the level of M-particle
states. Moreover,
a harmonic M-particle wave function can always
be written as a product of single particle oscillators
which leads to a Wigner density of the form

\be
\varrho
= \delta( \vec q_M-(\vec q_1+...+\vec q_{M-1}))\,
           \prod\limits_{j=1}^{M-1}\,8\,e^{-{|\vec t_j|^2}/{(x_j^0)^2}}\,
e^{-|\vec q_j|^2\,(x_j^0)^2}
\ee
The $\vec t_j=\sum\limits_i(\hat C)_{ji}\vec x_i$ are given by the
linear transformation  of
the original
cartesian coordinates $\hat C$.
The generalized relative momenta are                 
$\vec q_j=\sum\limits_k \left(C^{-1,+}\right)_{jk}\vec p_k$.
For complicated systems the transformation
$\hat C$ can contain particle masses and different coupling constants.
Our purpose has been a simple parametrization rather than
taking into account detailed information in the wave function as e.g.
Coulomb repulsion.
Therefore, we use
only one coupling constant $D$ for each cluster
and an equal mass of 
protons and neutrons $m$ which leads to a representation in so-called Jacobi
coordinates:
For t and $^3$He states the two relative coordinates are
$\vec t_1=\sqrt{3D/2}(\vec x_2-\vec x_1)$, 
$\vec t_2=\sqrt{2D}(\vec x_3-(\vec x_1+\vec x_2)/2)$ and 
 $x_0^1=x_0^2=(3D/2m)^{1/4}$ while  
for $^4$He states  
$\vec t_1=\sqrt{2D}(\vec x_2-\vec x_1)$, 
$\vec t_2=\sqrt{8D/3}(\vec x_3-(\vec x_1+\vec x_2)/2)$,
$\vec t_3=\sqrt{3D}(\vec x_4-(\vec x_1+\vec x_2+\vec x_3)/3)$ and 
$x_0^1=x_0^2=x_0^3=(2D/m)^{1/4}$.
The coupling constants are adjusted
to the mean square charge radii of the diverse cluster states
(see Table 1 and \cite{di145,di70,di71} )

\section{Baryonic Mean Fields}
At nuclear ground state density the nuclear mean field may be decomposed 
into two large pieces: an attractive scalar field provided by 
the quark condensate and/or correlated two-pion exchange (the $\sigma$ field),
and a repulsive vector potential (the $\omega$ field) \cite{4}
which is in accord with Dirac phenomenology for optical potential
calculations \cite{pa1}-\cite{pa4} 
in p+A studies and QCD sum-rule estimates \cite{7}.
Not much is known about the strength of the mean fields at large
net baryon densities and temperatures predicted 
in all present transport approachs for the ultrarelativistic
regime. It is expected that the momentum dependence 
\cite{di129,pa1}, the excitation into 
resonances \cite{25} and the transition to
quark matter \cite{QGP} will play a crucial role for the created mean
fields.
Several new ideas are currently under active investigation:
Medium properties of hadrons (e.g. of the $\omega $ meson
which is responsible for  vector repulsion \cite{20})
or quark and gluon condensates, 
which break the approximate scale and
chiral symmetry of  the QCD Lagrangian in the vacuum,
could modify the scalar field essentially \cite{QGP,7}.

In the following we demonstrate the sensitivity
of flow observables to mean fields by comparing
two schematic cases: 
In the first case the potentials are switched off
(i.e. the $cascade$  $mode$ is used). 
The second scenario uses potential type interactions which define
effective baryon masses in a medium
\cite{9} 
\be
\label{masss}
 p_i^2-m_i^2-V_i=0 
\ee
and thus simulate the effect of mean fields.
Here
\begin{equation}
(2m_N)^{-1} V_i
=+{1\over2} 
 \sum_{j,j\not =i} \alpha_{ij} 
    \left({\rho_{ij}\over\rho_0}\right) 
     + {\beta\over \gamma+1} 
        \left[ \sum_{j,j\not =i}  
            \left({\rho_{ij}\over\rho_0}\right)\right]^\gamma  
\end{equation}
with $\rho_{ij}$ a Gaussian of the CMS distance vector normalized to one,
$\rho_0$ ground state matter density and $\alpha=-0.4356$GeV, 
$\beta=0.385$GeV, $\gamma=7/6$ parameters which are adjusted
to the saturation properties of nuclear matter
(binding energy and compressibility). The parameter fit was done
by taking the expectation value of the total energy
per nucleon for ideal gas (plane) wave-functions 
taking into account antisymmetrization.
A Hamiltonian can be constructed which conserves  the
 effective mass shell constraints of Eq. \ref{masss} \cite{9}
and is employed for the dynamical evolution in RQMD.
For the presented calculations  we have chosen to
let the particles interact at equal times in the 
global equal-speed system of projectile and target.
    
As has been stated in \cite{10}, the experimental data for
nucleus-nucleus reactions at 10-15AGeV  indicate more 
repulsion than just given by a pure density dependence as in
Eq. (3). 
In fact, it was found that the quasi-potentials do not affect
the final distributions at all, if their strength is
below a `critical threshold'.
This additional repulsion is probably caused by
 the
 momentum dependence of nuclear forces. 
In order to explore the possible role of mean fields, in
Ref. \cite{10} we hardened the density dependence of the potentials
until we achieved agreement with experimental proton  spectra
which were available at that time.
The attractive 2-body force in the
$\Delta\Delta$ and NB$^*$ channel have been switched off
-- $\alpha_{\Delta\Delta}=\alpha_{\rm N\rm B^*}=0$ -- (thus explaining
the index pair ($ij$) in Eq. (3)).
Here,   we use the same potentials. We note that this approach
has predictive power for the cluster spectra, because the
strength of the potentials and thus the amount of flow
has been fixed from the proton spectra alone.
   
  What does the
 existence of  a `critical threshold' above of which only the
  collective forces `win' against the
   randomization of the motion by stochastic collisions
   and decays  mean? It
  indicates that the  mean field effect cannot be isolated from
   the other -- stochastic -- interactions which are present
   in the system.
  For instance,  the initial baryon density is essentially
   fixed solely from the stochastic interactions, because
   the degree of energy degradation which the ingoing nucleons
   experience is mostly determined by the
   multiple collisions with nucleons from the
   other ingoing nucleus and with secondaries.
  More recent versions of RQMD 
    contain 
   somewhat more realistic interactions
   than used here for the presented calculations
   (see Ref. \cite{ropes}
  for a discussion). For instance, the 
    assumption of istropic heavy baryon resonance decay
   leads to an overestimation of the nuclear stopping  power
   as it was noted recently in Ref. \cite{ropes}.
    A detailed study of the interplay between 
     the effect of collisions and mean fields based on
    RQMD is currently undertaken by one of us \cite{HSNEW}. 
  On the other side, we do not expect that our qualitative
  conclusions about the effect of mean fields
   on cluster flow will be reversed by more realistic
  calculations. In fact, a  smaller initial baryon density  means
   that the mean fields have to get stronger to achieve the same
  amount of collective flow.

\section{Results and Discussion}
\begin{figure}
\vbox{\hbox to\hsize{\hfil
 \epsfxsize=4.0truein\epsffile[80 30 583 775]{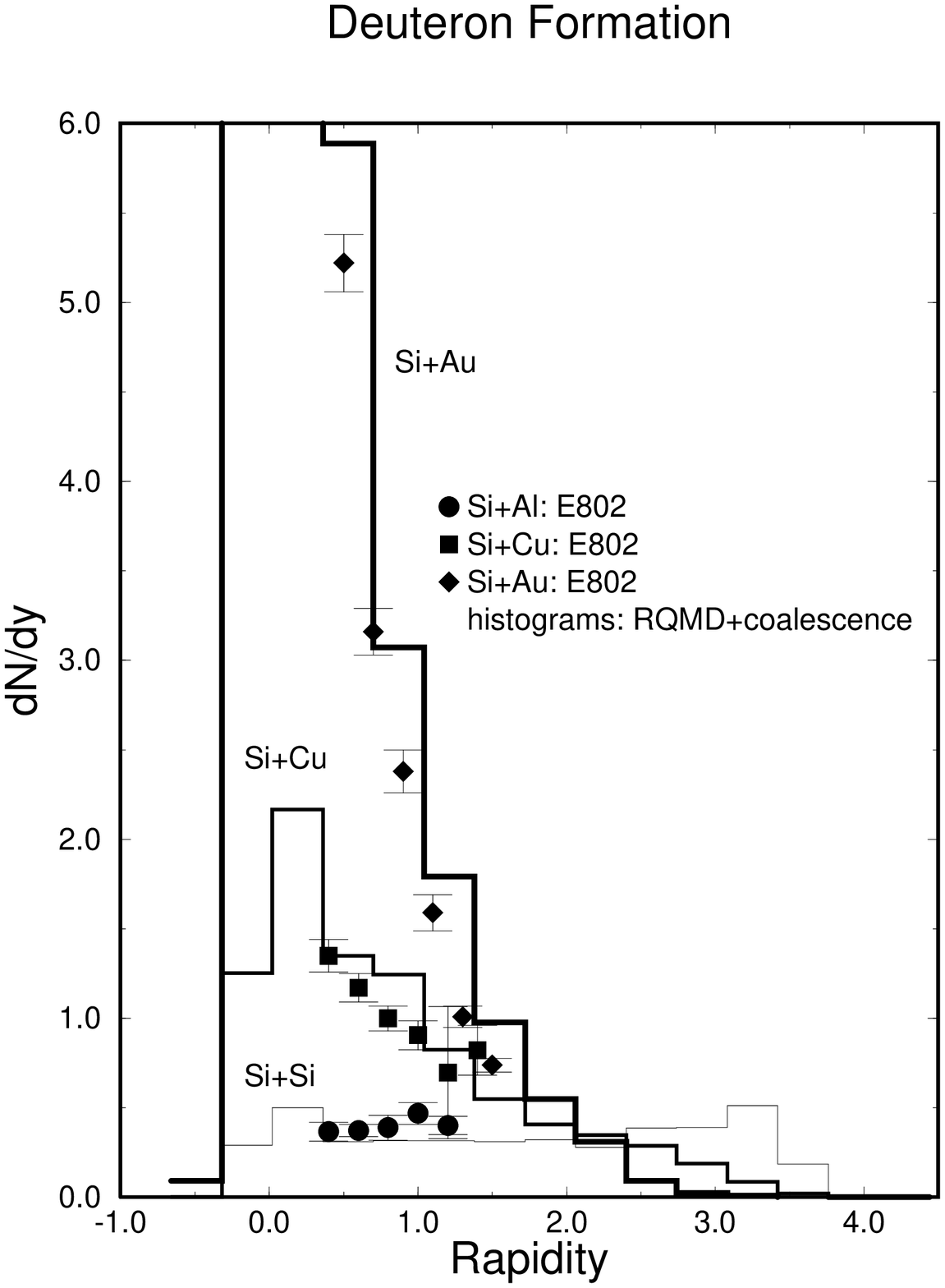}
 \hfil}}
\caption[] {
Rapidity distributions of deuterons in
Si+Si (b$<1$fm), Si+Cu (b$<1.5$fm) and Si+Au (b$<3$fm)
reactions at 14.6AGeV calculated from RQMD simulations including
potential interactions for baryons (solid histograms).
The symbols show E802-data from Ref. \cite{f12} for
central Si+Al, Cu and Au reactions. Note that the data
have been extrapolated in $m_t$ and contain $\approx 15\%$
systematic uncertainty.
           }
\label{Fig:deutdndy}
\end{figure}

\begin{figure}
\vbox{\hbox to\hsize{\hfil
 \epsfxsize=4.1truein\epsffile[80 30 583 775]{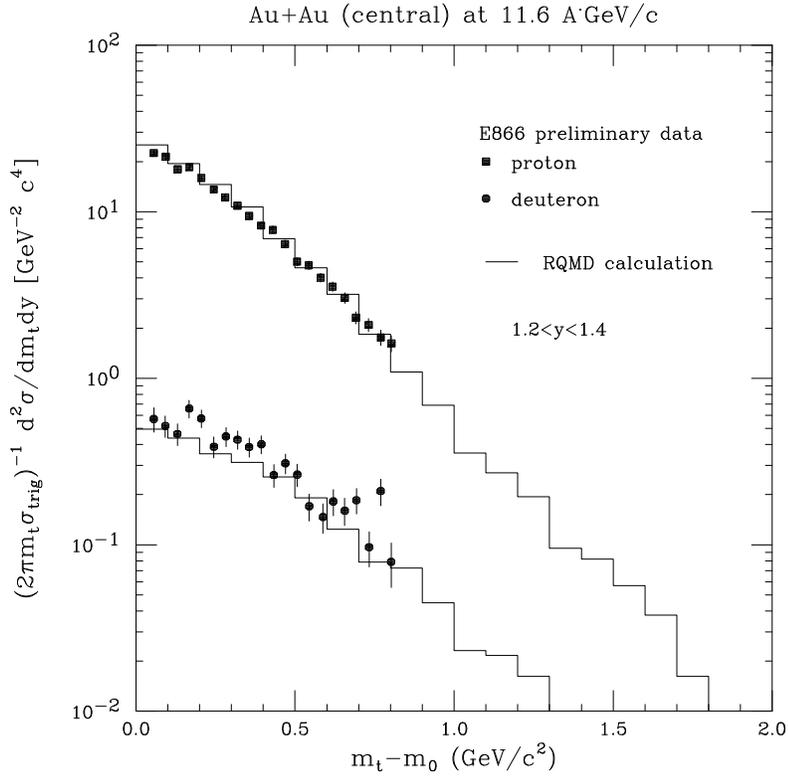}
 \hfil}}
\caption[]{
Transverse mass spectra for protons and deuterons
in central Au(11.6AGeV)Au reactions at $y_{LAB}=1.3$.
RQMD-simulations including potential interactions for baryons 
(histograms) are
compared with preliminary E866-data (symbols) \cite{bejin}. 
          }
\label{Fig:dmtcompare}
\end{figure}

The production of clusters has recently been measured and analyzed
for central and peripheral reactions p+A, Si+A and Au+Au at AGS-energies
\cite{di125,f12,2,AGSclust} and S+A at 200AGeV \cite{gillo}.
Comparisons between coalescence calculations for deuterons
at $p_t=0$ and measurements for pA reactions have also been discussed  
in \cite{di125,jamie}. We will first show that the
phase space coalescence in combination with the RQMD freeze-out
describes absolute values and momentum distributions of deuterons
in accordance with measurements for various nucleus-nucleus collisions.
In Fig. \ref{Fig:deutdndy} we compare our recently published
results for deuterons in central Si+A reactions 
at 14.6AGeV \cite{di18} with E802 data \cite{f12}.
In Fig. \ref{Fig:dmtcompare} calculations for transverse mass spectra
of protons and deuterons in the reaction Au(11.6AGeV)Au (b$<$3fm) 
at rapidity $y=1.3$ are compared to preliminary E866-data \cite{bejin}
for central events. 
A  comparison between calculations and data requires
proper event selection according to experimental trigger conditions
and acceptance corrections for the theoretical calculations 
\cite{di52}. On the level of the systematic errors in the 
measurements ($\approx$ 15\%, 
\cite{di52,f12}), however, we find good agreement -- even for the
strong slope parameter  splitting between protons and deuterons
in massive reactions.

\subsection{Transverse Expansion and Cluster Flow}
The formation of transverse blast waves was first proposed
in Ref. \cite{siemens} where  pion and proton 
transverse momentum spectra around 1AGeV incident beam energy were
analyzed.
The most prominent observables are the characteristic
shoulder-arm shape and different apparent temperatures for particles
with different mass. These are caused by the overlay of rather small 
local momenta and collective motion which have been produced during
the expansion phase of the reaction.
Several investigations followed  in the
low energy regime \cite{ogloblin,Daniel,eostrans,konopka} 
as well as for  ultrarelativistic
nucleus-nucleus reactions \cite{15,di59,stachel,greek,di18,clscaling}.

In the following we will discuss the momentum spectra
of nuclear clusters which show a strong dependence on such 
phase space correlations. In particular heavy clusters 
like $^4$He can serve as a very promising tool 
to determine the  phase space picture as e.g. provided by the 
microscopic transport calculations.
The strongest flow and
mean field effects are achieved in
central reactions Au(11.6AGeV)Au. 
Variations with the reaction size and life time of the
high density zone are studied by  comparison with  results for 
Si+Si reactions at 14.6AGeV. All results presented here
have been calculated for central impact parameters 
(Au+Au: b$<$3fm, Si+Si: b$<$1fm). 
The weak decay of hadrons after the freeze-out has been suppressed.

\begin{figure}
\vbox{\hbox to\hsize{\hfil
 \epsfxsize=4.1truein\epsffile[80 30 583 775]{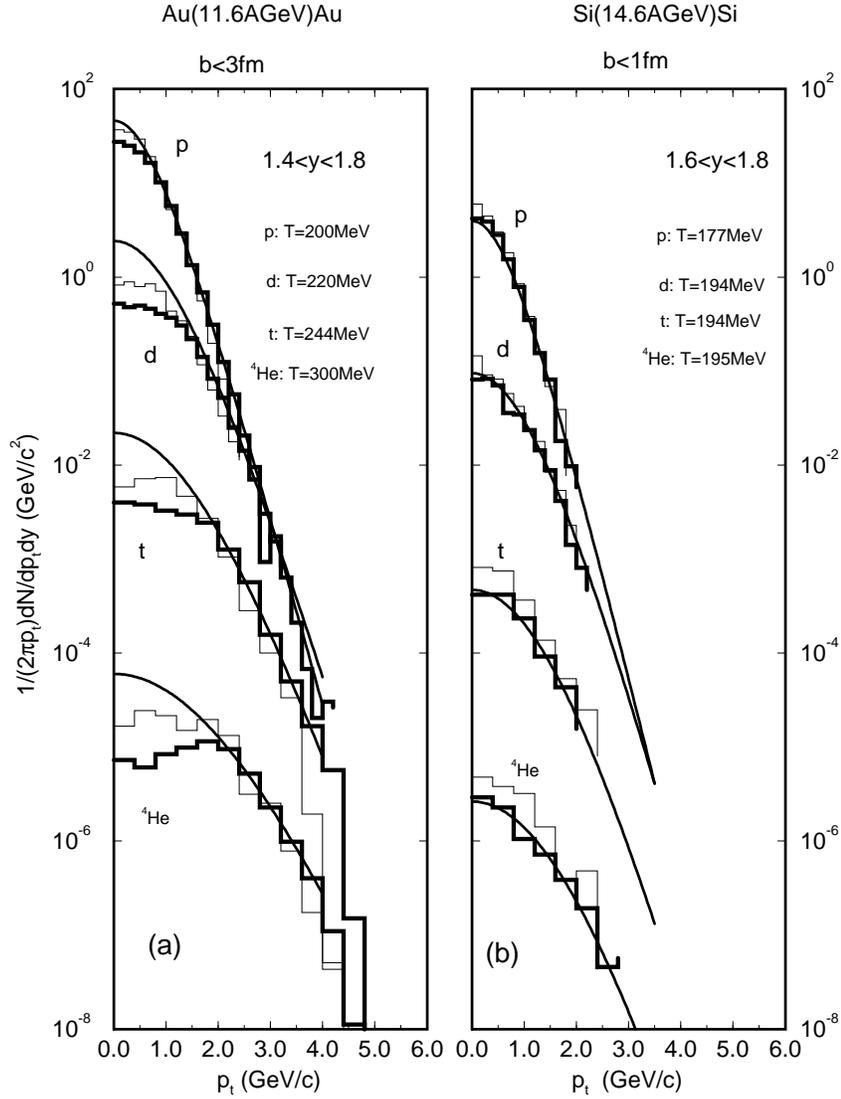}
 \hfil}}
\caption[]{
Transverse momemtum spectra for p,d,t and $^4$He
in Au(11.6AGeV)Au, b$<3$fm (a) and
Si(14.6AGeV)Si, b$<1$fm (b) at central rapidities.
Calculations including baryon potentials
(bold solid histograms) are compared with cascade simulations
(thin solid histograms). The smooth solid lines show
Boltzmann parametrizations adjusted to the high momentum
part of the spectra 
(see text) in calculations with potential interaction. 
          }
\label{Fig:dndptnew}
\end{figure}

Rapidity distributions 
and transverse momentum spectra
of p, d, t and $^4$He cluster are shown in
Figs. \ref{Fig:dndy4} and \ref{Fig:dndptnew}
for central Au+Au (11.6AGeV) and
Si+Si (14.6AGeV) reactions.
The figures contain cascade (solid histograms)
and potential calculations (bold solid histograms).
The solid lines in Fig. \ref{Fig:dndptnew} show
Boltzmann distributions with temperature parameters adjusted to
fit the transverse spectra for $p_t>$2GeV/c (Au+Au) and
$p_t>$1GeV/c (Si+Si) 
as calculated with baryon potential-interaction.

In Au+Au collisions all rapidity distributions peak at midrapidity 
indicating strong stopping in accordance with 
earlier predictions and preliminary data for protons
\cite{Keitz,di18}. 
The transverse distributions have
a strong shoulder-arm shape which deviates markedly from
distributions expected from a purely thermalized fireball. 
The shoulder-arm shape becomes most prominent for 
heavy clusters. 
For $^4$He clusters a peak even appears at finite $p_t$. 
The  high momentum tail of the transverse spectra 
exhibit different 'apparent' temperatures for clusters with different 
mass while a thermal system would
predict similar slope parameters \cite{di131}.
Note that the extracted temperature values depend strongly on the $p_t$-cut
choosen. The absolute values extracted by exponential parametrizations 
always lead to additional systematic errors in the
absolute yields 
according to our calculation overestimating the cluster
yields substantially. 
Therefore, comparisons of rapidity distributions between 
calculations and extrapolated data 
must be done very carefully.

In contrast, all rapidity distributions
are essentially flat for the light system Si+Si.
The transverse spectra are in good agreement with a thermal
fit and show temperature parameters which are almost equal for all states.

\begin{figure}
\vbox{\hbox to\hsize{\hfil
 \epsfxsize=4.1truein\epsffile[80 30 583 775]{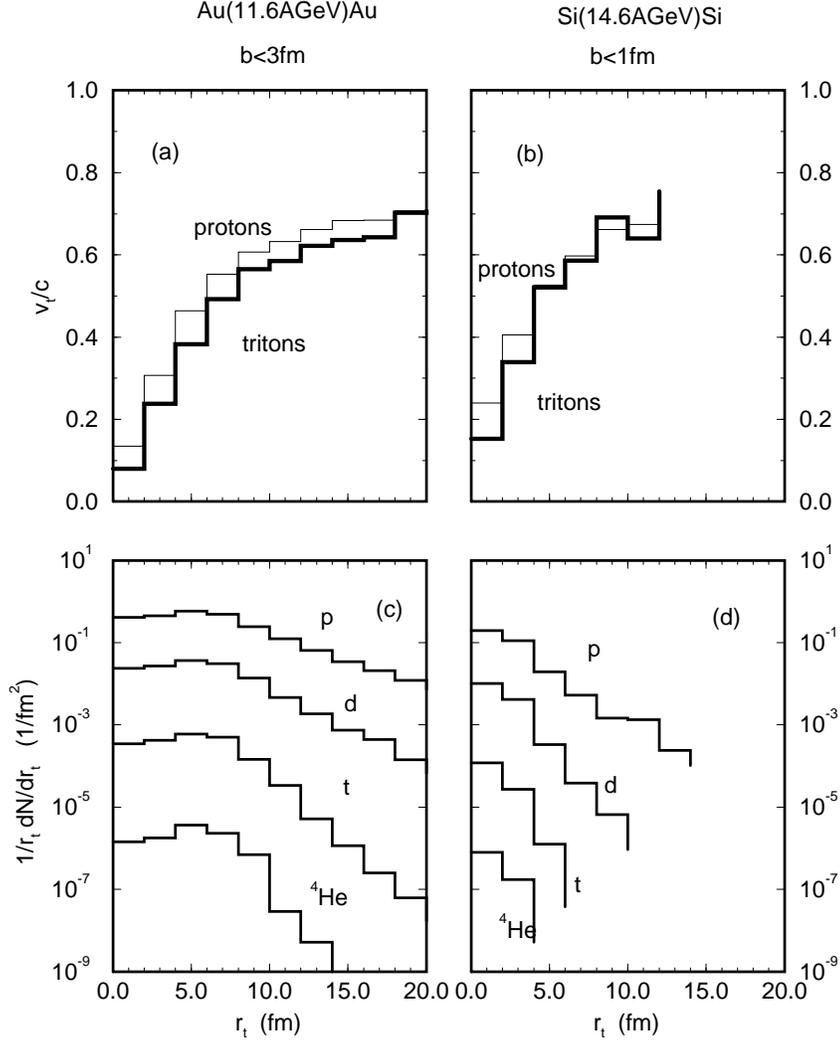}
 \hfil}}
\caption[]{
Freeze-out profiles of protons and tritons 
 in RQMD-calculations including baryon potentials for
 Au(11.6AGeV)Au, b$<3$fm (a,c)
and Si(14.6AGeV)Si, b$<1$fm (b,d) at central rapidities. 
The upper part shows
transverse velocity profiles for protons and tritons.
The lower part shows the distributions of transverse
freeze-out densities of p, d, t and $^4$He.
          }
\label{Fig:profiles1}
\end{figure}

The characteristic deviations from  thermal distributions
are caused by 
strong transverse expansion and collective flow
particularly in massive reactions like Au+Au. 
The flow-correlations at the microscopic freeze-out are 
shown in Fig. \ref{Fig:profiles1} which contains calculations for 
the freeze-out velocities and  density profiles
of protons, deuterons, tritons and $^4$He.
The velocity profiles for all clusters are similar.
They exhibit a convex shape and saturate at $\approx$0.7c for both 
reactions. 
In the massive system Au+Au the freeze-out densities have
a complicated shape which peakes around 5fm. Most of the nucleons
freeze out at larger distance. This is indicated by an average freeze-out 
radius of $\approx 10$fm (see Table 2).
The strong transverse expansion in Au+Au collisions is caused by 
the considerable baryon stopping and the pile up of
high particle densities near to the reaction center. 
During the expansion phase comoving particles undergo 
frequent collisions transporting the system collectively sidewards 
until the flow-induced pressure pushes them into the vacuum.  
Hence, the suppression of particle emission 
at $r_t\to0$ is basically caused by the dynamical expansion:  
many nucleons are transported through the medium before they reach
the 'surface' and freeze out. 
Table 2  gives the average values for freeze-out radii, 
collective flow velocities and transverse momenta
of particles in the central rapidity region.
Note that the average transverse 'velocities' $<p_t>/A$, the transverse
flow velocities and source radii decrease with
increasing A and saturate for cluster masses $A\ge3$.
The  relative  suppression of cluster formation at the
'surface' ($r_t>6$fm) contradicts the  present fireball analyses which
assume a common density and velocity profile for all particles.

Clusters are clearly dominated by the collective flow components
in the final phase space distribution:
The collective transverse velocities of heavier mass clusters
already provide  most of the total transverse momentum ($\approx$80\%).
Therefore, the freeze-out density and collective 
velocity profiles 
determine the final spectra almost exclusively: 

It is convenient to
approximate the collective velocity profile by
$\beta_t$$=$$A(r_t-r_0)^B$
in order to demonstrate the effect of the
interwining of collective velocities and freeze-out probabilities
on the transverse spectra. This parametrization
leads to a transverse momentum spectrum of the form
\begin{equation}
{{\rm d}N\over p_t{\rm d}p_t}
  = {m_0^2\over B A^{2/B}}
  { (m_t^2-m_0^2)^{ {(1-B)/B} }\over
        (m_t^2)^{ {(1+B)/B} }}
  \rho(r_t) \quad;\quad m_t:=\sqrt{m_0^2+p_t^2}
\end{equation}
where $\rho$$=$$1/(r_t-r_0){\rm d}N/{\rm d}r_t$.
The only quantities which  determine such a purely flow-induced spectrum are
the shape of the velocity profile
defined by the parameter $B$ and the density distribution $\rho$. 
Assuming a box density profile
the spectra have a convex shape and maximum
at finite $p_t$  only for $B$$<$$1$ -- consistent with the RQMD results.
A quadratic $r_t$-dependence ($B$$=$$2$) would instead yield an overall
concave spectrum, in particular diverging at $p_t\to0$.

The flattening of the transverse spectra
at low $p_t$-values is due to the suppression
of clusters 
in the 
very central region of the reaction ($r_t\to0$).
The different apparent temperatures at high $p_t$-values
are caused by the strong weight of large flow velocities
for $r_t>6$fm. The peak/shoulder in the transverse spectra, however,
appears approximately at $p_t/A\approx <\beta_t>$ and directly 
measures the strength of the transverse flow 
at the position
where most of the clusters at central rapidities freeze out
($r_t\approx5-7$fm). 
Note that it is not possible to describe the 
transverse spectra with one single temperature and collective 
flow velocity in contrast to what has been claimed for reactions 
in the 1GeV/n regime \cite{eostrans,konopka}.
 
Si+Si collisions do not provide a 
transverse expansion 
comparable to massive reactions
(Fig. \ref{Fig:profiles1}, Table 2).
Most of the 
clusters are emitted close to the beam axis
where the transverse flow velocities are small.   
In fact, the 'surface suppression' 
acts more strongly  in the case of the
smaller system, 
in accordance with the 
larger surface to volume ratio.
Note that 
here 
the average transverse flow velocity of $^4$He states
is approximately a factor of two smaller than for protons (Table 2). 
Transverse flow is 
nearly invisible due to 
large  'local' momentum components.
This sampling of clusters at smaller distances from
the beam axis
explains why the transverse flow features are 
mostly
invisible,
although the velocity profiles for cluster states in Fig. \ref{Fig:profiles1} 
-- the 'collective' position-momentum correlations -- 
are almost equal for Si+Si and Au+Au collisions.

The role of the shapes of collective velocity and
density profiles has been  the subject of  much previous debate
\cite{ogloblin,di18,eostrans,15,di59,konopka}.
In particular the low-$p_t$ pion enhancement 
and the spectra of protons and deuterons
have been interpreted in terms of transverse flow
with the assumption of an expanding thermal fireball \cite{15,di59}.
A similar picture has been used to extract 'local temperatures'
and chemical potentials exploiting the final particle ratios
\cite{stachel,greek}. 
We wish to discuss this issue in more detail because in those  
analyses the freeze-out profiles used  differ significantly
from the results of the transport calculations presented here.

In Ref.  \cite{15} a value for $B=2$  in combination with a box-shaped 
density profile was used  to explain the low-pt pion enhancement
i.e. a concavely curved $p_t$-spectrum. As a consequence of these
assumptions  proton and deuteron spectra show the same behaviour,
in particular for $p_t\to0$.
The microscopic calculation, however, contradicts this picture and
shows profiles which are compatible 
with $B\approx0.5$ and a non-trivial position geometry.
The concave curvature of transverse spectra 
in the transport calculations has been
confirmed by recent data (see protons and deuterons in Si+A reactions
at 14.6AGeV \cite{f12} and the results for Au(11.6AGeV)Au
in Fig. \ref{Fig:dmtcompare}).
In this sense, neither the assumption $\rho=const$ nor
a  shape of the freeze-out profile according to $B=2$
--as used in \cite{15}-- can be justified.
The main reason for the misleading results in \cite{15} is probably the 
misinterpretation of concavely shaped pion spectra.  
Pions are strongly influenced by the final decays of resonances 
such as $\Delta,\rho,B^*$ (see \cite{di409} and Refs. therein). 
The alternative prediction that the low-$p_t$
pion excess at AGS energy comes from $\Delta$-resonances 
\cite{BSW,8}
 has been confirmed by experimental reconstruction of
 the p$\pi$ invariant masses which show a strong $\Delta$-signal,
  in agreement with RQMD \cite{E814}.
Furthermore, the early preliminary data for protons used in \cite{15} 
were limited in acceptance ($m_t-m_0>200$MeV). They 
excluded those regions where most of the shoulder-arm effect appears
and were -- within the error bars -- also consistent with concavely 
shaped distributions for protons and deuterons.

\subsection{Directed Flow}
\begin{figure}
\vbox{\hbox to\hsize{\hfil
 \epsfxsize=4.1truein\epsffile[80 30 583 775]{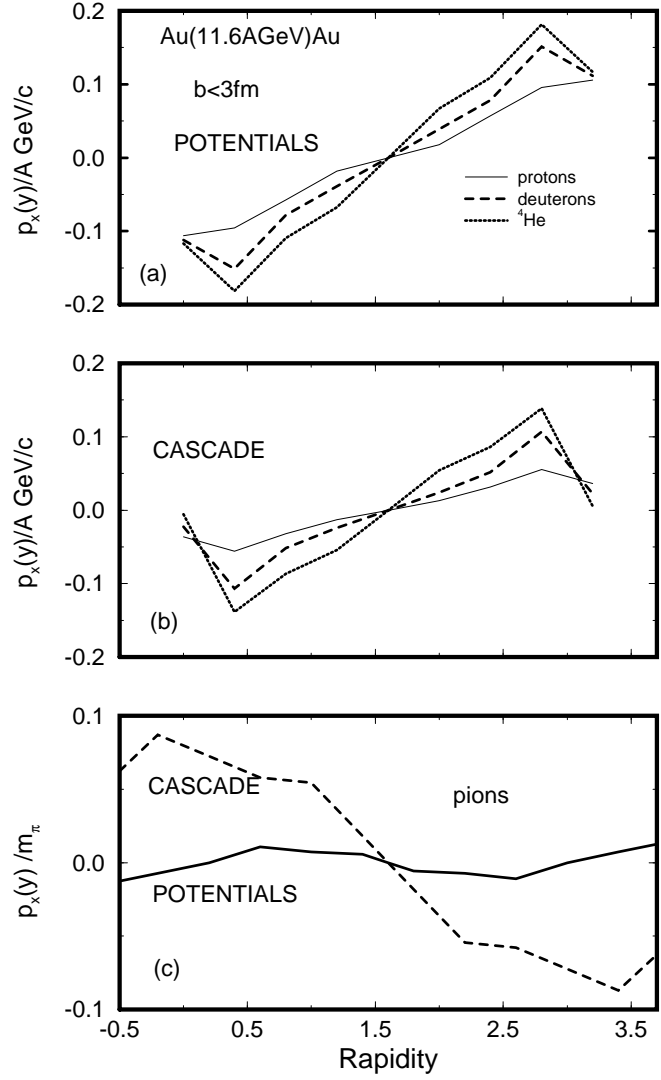}
 \hfil}}
\caption[]{
$p_x(y)/A$-correlations for p, d, $^4$He (a,b) and pions (c) in
central Au(11.6AGeV)Au (b$<3$fm) reactions.
The figure shows a factor of two increase of the cluster flow
if baryon potentials are included. 
The additional rotation of the
event plane due to the potentials leads to an apparent
vanishing of the pion flow in the laboratory system which
is, however, still pronounced in the principal axis system 
of the rotated (baryon) event. 
          }
\label{Fig:pxy_auau4}
\end{figure}

Besides the characteristic signals in the inclusive spectra,
the correlation between rapidity and directed transverse 
momentum $p_x(y)$ in Fig. \ref{Fig:pxy_auau4} 
is another indicator of a  non-trivial event geometry.
This observable is well known as the  nuclear ``bounce-off''
discovered first at the Bevalac \cite{Gutbrod}.
The averaged transverse velocity $p_x/A$ in Fig. 
\ref{Fig:pxy_auau4} is defined by the averaged transverse momentum 
per cluster nucleon projected on the theoretical reaction plane
for particles within  a certain rapidity interval $\Delta y$: 
\be
p_x(y_0)/A:= \left.\left\langle{1\over N} \sum_i {\hat e_x\cdot\vec p_i/A}
         \right\rangle\right|_{|y-y_0|<\Delta y/2}
\ee
$\hat e_x$ is the unity vector which points perpendicular to the
beam axis into the impact parameter plane. 
$\langle...\rangle$ denotes the final event averaging.
In the case of strong longitudinal and transverse flow 
contributions in the final source this quantity reflects the collective 
sideward flow of matter predicted by hydrodynamics 
\cite{21,Strott} and microscopic models \cite{21,Keitz,Strott}. 

\begin{figure}
\vbox{\hbox to\hsize{\hfil
 \epsfxsize=4.1truein\epsffile[80 30 583 775]{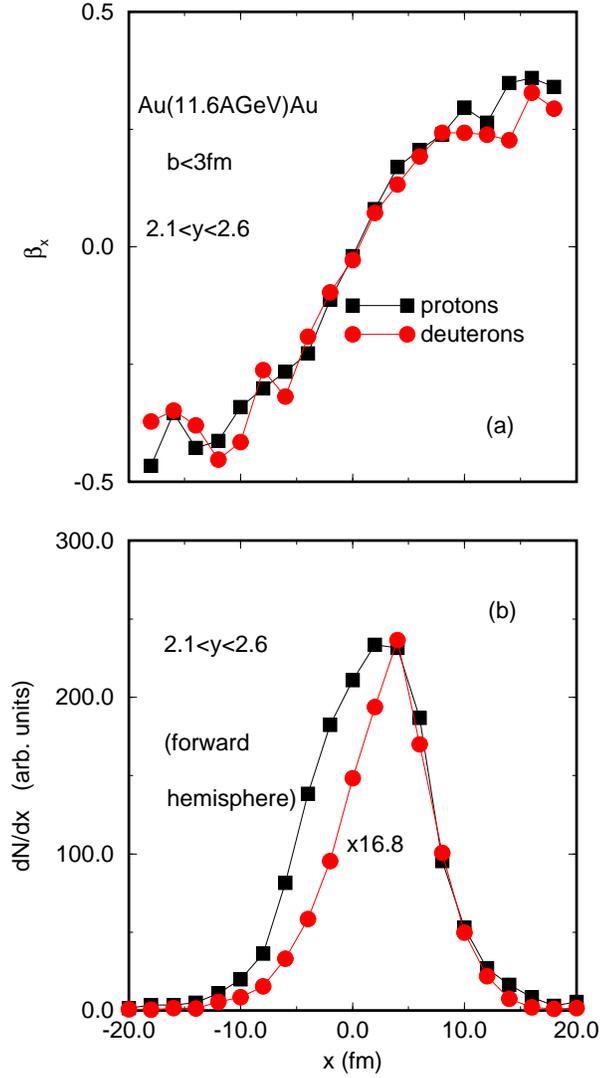}
 \hfil}}
\caption[]{
In-plane freeze-out velocity (a) and density profiles (b) of
protons and deuterons in central Au(11.6AGeV)Au reactions (b$<3$fm).
Selected are particles
in the forward ($2.1<y_{\rm Lab}<2.6$) hemisphere.
$x$ denotes the projection of the freeze-out position
onto the theoretical reaction plane. The deuteron density
is scaled by a factor 16.8 to demonstrate the 
increase in the transverse freeze-out distances from the 
beam axis between protons to deuterons. 
          }
\label{Fig:auxprof}
\end{figure}

Clusters exhibit larger $p_x(y)/A$ values than nucleons
although the division by $A$ excludes the trivial effect of 
the momentum scaling with mass $p_A/A\simeq p_N$ at equal velocity. 
This stronger correlation for cluster states is well known from
Au+Au reactions in the 1GeV/n energy regime \cite{eosflow,experiment}.
The reason for this behaviour is demonstrated in 
Fig. \ref{Fig:auxprof} which shows
profiles for the collective in-plane velocity $\beta_x=\hat e_x\cdot\vec p/E$
and freeze-out density of protons and deuterons.
Only particles in the forward hemisphere $2.1<y<2.6$ are taken into account.
The profiles are drawn  as a function of the transverse distance
to the beam axis taken in the original nucleon-nucleon cms and projected 
onto the theoretical reaction plane $x:=\vec x\cdot \hat e_x$. 

The densities for deuterons are scaled by an arbitrary factor to exhibit
the qualitative difference between nucleons and deuterons:
In contrast to the average values, the 'local' velocities of 
protons and deuterons  are equal.
The density distribution of the deuterons, however, exhibits a shift 
towards the outward regions as compared to protons.
This suppression of cluster formation near to the original beam axis 
is caused by higher relative momenta for nucleons. 
Therefore, the high transverse in-plane velocities are more strongly
weighted in the case of cluster formation which leads to higher $average$ 
velocities.  
Note the qualitative difference between this increase 
in the reaction plane in contrast to decreasing values for 
$<p_t>/$A at central rapidities.

\subsection{Mean Field Dependences}
\begin{figure}
\vbox{\hbox to\hsize{\hfil
 \epsfxsize=4.1truein\epsffile[80 30 583 775]{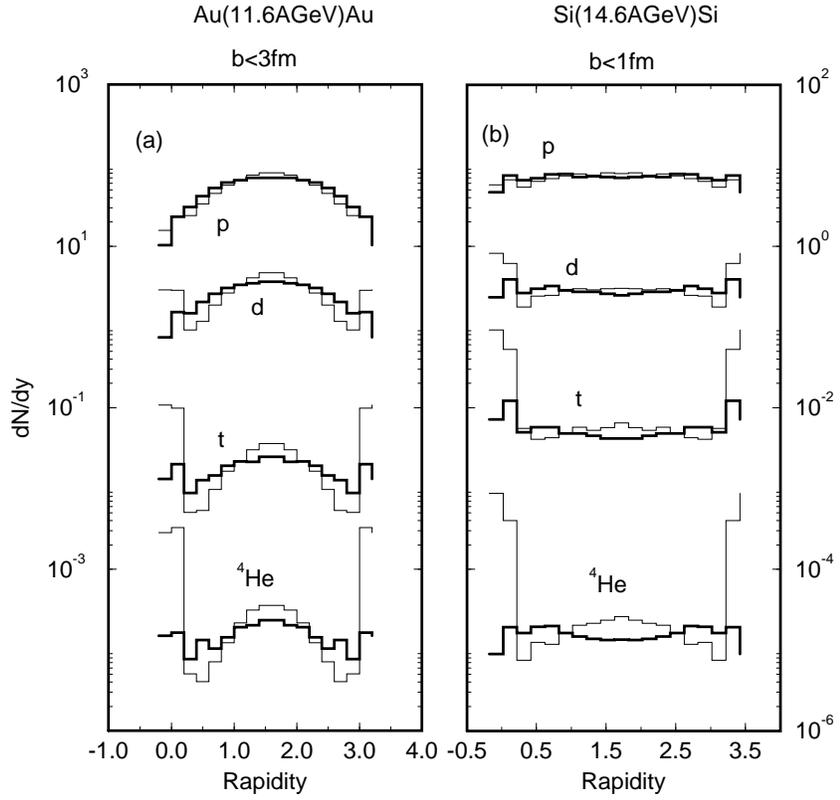}
 \hfil}}
\caption[]{
Rapidity distributions for p,d,t and $^4$He in
Au(11.6AGeV)Au, b$<3$fm (a) and
Si(14.6AGeV)Si, b$<1$fm (b).
Calculations with baryon potentials are denoted by bold solid
histograms. Cascade calculations are shown by thin solid histograms.
The inclusion of potentials at high baryon densities leads to
stronger longitudinal expansion in both systems. 
          }
\label{Fig:dndy4}
\end{figure}

The results in calculations for Au+Au 
show higher longitudinal (Fig. \ref{Fig:dndy4}) and transverse momenta
(Fig. \ref{Fig:dndptnew}, Table 2) caused by the 
additional pressure which is built up by the repulsive mean fields 
at high baryon density (up to 8$\rho_0$ is achieved \cite{Keitz}).
Note that the region of highest compression 
($\rho/\rho_0$$>$$3$) is large (V$\simeq$several hundred fm$^3$) 
and contains up to 60\% baryons in resonance states \cite{di409}. 

\begin{figure}
\vbox{\hbox to\hsize{\hfil
 \epsfxsize=4.1truein\epsffile[80 30 583 775]{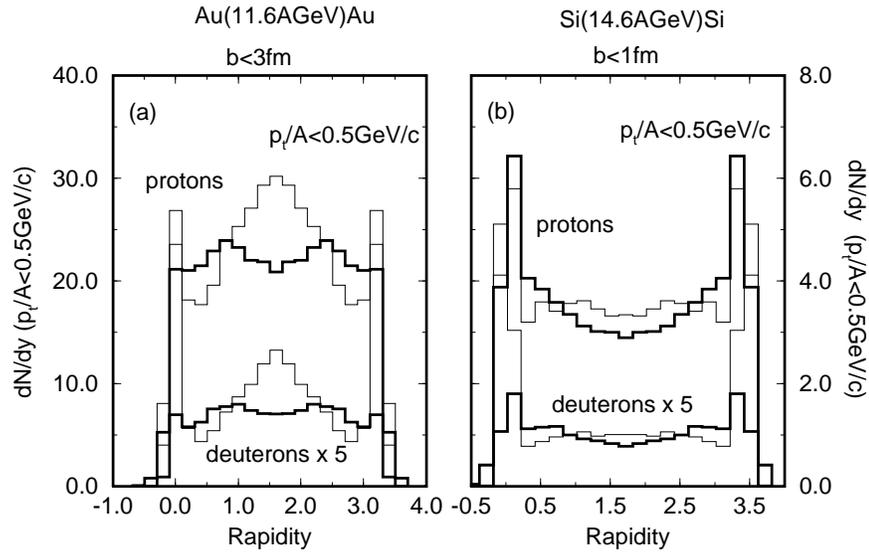}
 \hfil}}
\caption[]{
Comparison of dN/dy-distributions of protons and deuterons
including a 
transverse momentum cut $p_t/A<0.5$AGeV in calculations
with (bold solid histograms)
and without baryon potentials (thin solid histograms)
for central Au(11.6GeV)Au (a) and Si(14.6AGeV)Si (b) reactions.
          }
\label{Fig:dndy_ptcut1}
\end{figure}

\begin{figure}
\vbox{\hbox to\hsize{\hfil
 \epsfxsize=4.1truein\epsffile[80 30 583 775]{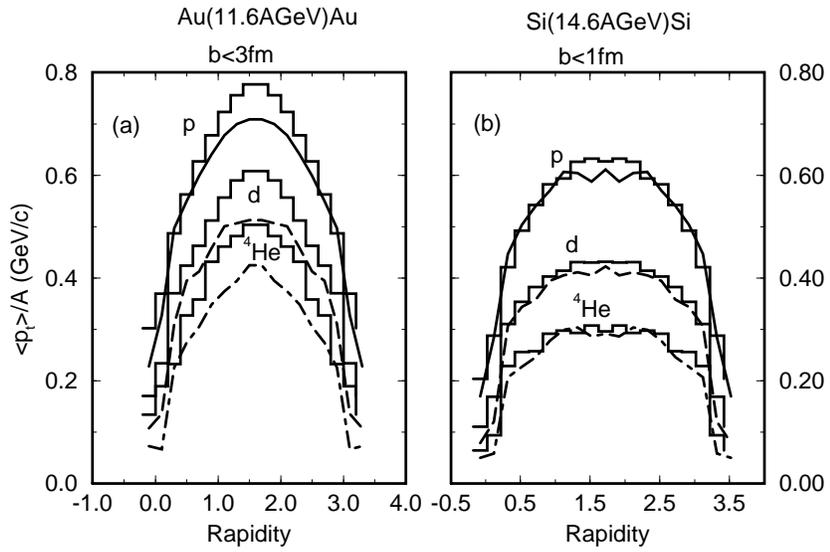}
 \hfil}}
\caption[]{
Average transverse momenta $<p_t>(y)$ of p, d, $^4$He
in calculations with (histograms) and without (lines) baryon potentials
for central Au(11.6GeV)Au (a) and Si(14.6AGeV)Si (b) reactions.
          }
\label{Fig:pty}
\end{figure}

The difference between potential and cascade calculations is 
largest in the low-$p_t$ part of the spectra (Fig. \ref{Fig:dndptnew}).
For heavier clusters the distributions close to $p_t=0$ change by up to a 
factor of three.
Nuclear matter at midrapidities is mostly affected by the mean field 
contributions at high baryon densities: 
Figs. \ref{Fig:dndy_ptcut1} and \ref{Fig:pty} show the 
rapidity dependence of the proton and deuteron yields at
low transverse momenta -- here defined by a 
cut in transverse momentum $p_t/A<0.5$GeV --
and the average transverse momenta of p, d and $^4$He.
While cascade calculations exhibit
a clear peak in the $dN/dy$ spectra, the calculations including
potentials show a dip even for central events.
This is due to the additional longitudinal expansion
caused by the baryon potential interaction. 
The potentials
also change the average transverse momenta by $<$20\%.

For the smaller system Si+Si the average transverse momenta
change by less than 3\%
including potential-interaction. 
In earlier work we have shown that the life time of the reaction
is  not  long enough to establish a thermalized
high density phase \cite{Keitz}. 
The 'transverse communication' is 
in light systems
much smaller and does not allow
for a considerable transverse push due to the mean fields,
although,  potentials also lead  to qualitative changes in the
distribution of the longitudinal momenta for the small system: 
The results in Fig. \ref{Fig:dndy4} show that cascade calculations exhibit 
a concave shaped spectrum which turns to a convex 
distribution if potentials are included. 
As in Au+Au reactions the yields of cluster states
are most sensitive: 
The dN/dy values for $^4$He states at midrapidity differ by almost
a factor of 2 between the cascade and the potential calculations.

The value of the flow-correlation $p_x(y)$ in central Au(11.6AGeV)Au 
collisions is roughly a factor 1.5-2 higher 
due to the additional sideward push 
of the mean fields (Fig. \ref{Fig:pxy_auau4}).
In recent work we have shown that pions and other produced 
hadrons (antikaons, antinucleons) show a characteristic
'anti-flow' \cite{di402} 
caused by scattering off spectator-like
matter. This behaviour has previously been discussed  
in reactions at 1AGeV \cite{di198,densflow,konopka}. 
Fig. \ref{Fig:pxy_auau4}  includes this
pionic anti-flow which also appears 
to be sensitive to the baryonic mean fields, too: 
While cascade calculations show 
sizeable $p_x/m$ values for $\pi$s, 
the inclusion of baryonic potentials leads to almost vanishing
$p_x(y)$-values in the laboratory frame. 
In the principal axis system, however,
the strong anticorrelation of pions to baryons is conserved.
In the work of Li et al. \cite{Li} the in-plane pion flow 
has been investigated in the framework of the
cascade model ART. In these calculations
the sign of the pionic flow is equal to the baryon flow 
in central events which is in qualitative difference to the RQMD results. 
Note the strong dependence of the in-plane
pion flow on different  absorption rates at high baryon densities
which has recently been analyzed 
for reactions at lower incident
beam energies with the QMD-model \cite{densflow}.

Both  results for central Au(11.6AGeV)Au collisions including 
baryon potentials (high in-plane flow for nucleons, small 
$p_x/m$ for pions)
are in quantitative agreement  with preliminary flow measurements
from E877 \cite{wessels} 
and E866.
The convex proton rapidity
distribution in Si+Si and the width of the distribution in 
 Au+Au reactions, including potential interactions, are also in
 accord with published \cite{f12} and preliminary \cite{2,di18}
data. 
Nevertheless, even potential calculations 
overestimate stopping and underestimate the transverse momentum
production at forward rapidities ($y>y_{NN}$) 
in asymmetric reactions like Si+Au \cite{di52}.
The results for deuterons in Fig. \ref{Fig:deutdndy} 
show the same trend.

\section{Summary and Outlook}
We have presented a phase space coalescence model            
for cluster formation 
with $A\ge 2$
in relativistic nucleus-nucleus collisions
using the Wigner-function method. 
The combination    with the
transport approach RQMD allows the calculation of cluster momentum 
distributions. Specifically, we have studied  
central Au(11.6\-AGeV)\-+Au and Si(14.6AGeV)+A reactions.
The microscopic model shows that  strong stopping 
results in observable collective behaviour
of the stopped baryon-rich matter.
Considerable flow ($<\beta>$$\approx$$0.5$ c) develops due
to the internal pressure of the dense matter.
The transverse expansion is most visible in the 
momentum spectra of nuclear clusters which deviate markedly from 
thermal distributions.
In central Au(11.6AGeV)Au collisions the transverse spectra exhibit
a strong shoulder-arm shape which is most prominent for
heavier mass clusters.

The 'apparent' temperatures at high transverse momenta,
which result from the overlay of rather small local momentum 
fluctuations and collective flow velocities,
increase with cluster mass.
Furthermore, a clear ``bounce-off'' event shape is seen in massive reactions
like Au(11.6AGeV)Au. The averaged transverse flow velocities 
in the reaction plane $<p_x>(y)/A$
are markedly larger for clusters
than for protons (by a factor of 2).  
Both the shoulder-arm shape and the large bounce-off signal for nuclear 
clusters are directly related to the freeze-out geometry and flow 
correlations. 
In contrast to the directed flow
our results show smaller freeze-out radii and
smaller average transverse 'velocities' $<p_t>/A$ 
for clusters at midrapidities
than for free protons. This result is in contradiction to simple
fireball and blast wave models.

The  cluster spectra and the in-plane flow may change markedly if baryon  
potential-interactions are included:
According to the calculations presented here
the yields of nuclear clusters decrease by 
up to a factor of three, 
at low $p_t$ and central
rapidities,
in the reaction Au(11.6AGeV)Au.
The average $p_t/A$ values are $\approx$15\% harder
in calculations with potentials than in the cascade mode.
The 
$p_x(y)$-correlation for nucleons and
nuclear clusters increases by
a factor of
1.5-2 while
the anticorrelated in-plane flow of pions vanishes.
For central Si(14.6AGeV)Si reactions the potentials play a negligible 
role in transverse direction, but affect the
proton and cluster rapidity spectra.
Cascade calculations exhibit 
concave spectra which turn into convex 
distributions if potentials are taken into account.
The absolute yield for $^4$He clusters changes by
almost a factor of two at midrapidity.
The strong sensitivity of nuclear clusters 
 to the collective flow
 encourage the  quantitative study of the   
 transient pressure in nucleus-nucleus collisions.

\section*{Acknowledgement}

This work has been supported by the BMFT, GSI, DFG and
the A.v.Humboldt-foundation. 
One of us (R.M.) wants to
thank the nuclear theory and experimental groups at BNL and SUNY 
for their 
kind hospitality
and useful discussions. 

%

\clearpage
\clearpage
\clearpage
\clearpage
\clearpage
\clearpage
\clearpage
\clearpage
\clearpage

\newpage

\begin{center}
\begin{tabular}{|l|c|c|}
\hline
\multicolumn{3}{|l|}{Table 1}\\
\hline
cluster & radius (fm) & D (GeV$^3$) \\
\hline
$^3{\rm H} $  & $1.7$   &   $0.27\times10^{-4}$ \\ 
$^3{\rm He}$  & $1.8$   &   $0.34\times10^{-4}$\\
$^4{\rm He}$  & $1.5$   &   $0.53\times10^{-4}$ \\
\hline
\end{tabular}
\end{center}

\vspace{1cm}

{\bf Tab. 1} Root mean square charge radii for nuclear clusters
with A=3,4 and the corresponding
coupling strength in the harmonic oscillator approach.

\newpage

\vspace{3mm}

\hspace{-8mm}
\begin{tabular}{|l|c|c|c|c|c|c||c|c|c|c|c|c|}
\hline
\multicolumn{13}{|l|}{Table 2}\\
\hline
                &   n  &   p  &   d   &   t   & $^3$He & $^4$He &
                    n  &   p  &   d   &   t   & $^3$He & $^4$He \\
\hline
\multicolumn{7}{|c||}{Au+Au cascade}&\multicolumn{6}{|c|}{Au+Au potentials}\\
\hline
$<\beta_t>$     & $.5$ & $.5$ & $.42$ & $.37$ & $.38$  & $.36$ &
      $.54$ & $.54$ & $.47$ & $.43$ & $.44$  & $.42$ \\
$\sqrt{<r_t^2>}$ (fm) &  9.5 &  9.7 &  7.9  &  6.6  &  6.2   &  6.0 &
  10.0 &  9.9 &  7.9  &  6.6  &  6.9   &  6.2 \\
$<p_t>/A $(GeV/c) & 0.71 & 0.71 & 0.51 & 0.44 & 0.44 & 0.43 &
                   0.77 & 0.78 & 0.61 & 0.51 & 0.53 & 0.50 \\ 
\hline
\multicolumn{7}{|c||}{Si+Si cascade}&\multicolumn{6}{|c|}{Si+Si potentials}\\
\hline
$<\beta_t>$     & $.39$ & $.39$ & $.32$ & $.26$ & $.26$  & $.24$ &
 $.41$ & $.40$ & $.31$ & $.24$ & $.26$  & $.23$ \\
$\sqrt{<r_t^2>}$ (fm)&  3.9 &  3.7 &  2.7  &  2.2  &  2.2   &  1.8 &
  3.9 &  3.9 &  2.7  &  2.2  &  2.3   &  2.1 \\ 
$<p_t>/A$ (GeV/c)& 0.62 & 0.61 & 0.42 & 0.32 & 0.33 & 0.29 &
                   0.64 & 0.63 & 0.43 & 0.33 & 0.34 & 0.30\\
\hline
\end{tabular}

\vspace{1cm}

{\bf Tab. 2} 
Average transverse freeze-out velocities $<\beta_t>$, freeze-out radii
$\sqrt{<r_t^2>}$ and average transverse momenta $<p_t>/A$
for nucleons and cluster with $A\le4$ in central
Au(11.6AGeV)Au and Si(14.6AGeV)Si reactions. The table contains calculations
with (r.h.s.) and without (l.h.s.) baryon potential-interaction. 
Only particles in the midrapidity region $1.4<y<1.8$ (Au+Au) and 
$1.6<y<1.8$ (Si+Si), respectively, are taken into account.


\begin{thebibliography}{99999999}
 \bibitem{19} J.B. Kogut, D.K. Sinclair and K.C. Wang: Phys. Lett.
 B163(1991)101
 \bibitem{20}
 U. Vogl and W.Weise: Prog. Part. Nucl. Phys. 27(1991)195;
 W. Weise: Nucl. Phys. A553(1993)59
\bibitem{7}
 X. Jin, T.D. Cohen, R.J. Furnstahl and D.K. Kriegel:
 Phys. Rev. C47(1993)2882 and refs. therein
\bibitem{2} M. Gonin for the E802/E866 collaboration:
 Nucl. Phys. A566(1993),601c 

\bibitem{E802}  
Hideki Hamagaki for the E-802/866 collaboration: 
Nucl. Phys. A566(1993),27c


\bibitem{E810} K.J. Foley for the E810 collaboration: 
Nucl. Phys. A544(1992)335c

\bibitem{E814} J. Stachel for the E-814/877 collaboration:
Nucl.Phys. A566(1993),183c

\bibitem{trenerg} J. Barette for the E-814/877 collaboration:
Nucl. Phys. A566(1993),411c



 \bibitem{9}
 H.Sorge, H. St\"ocker and W. Greiner: Ann. Phys. (NY) 192(1989)266;
                     Nucl. Phys. A498(1989)567c;
H. Sorge, A. v. Keitz, R. Mattiello, H. St\"ocker and W. Greiner:
                     Z. Phys. C47 (1990)629

 \bibitem{8}
H. Sorge, R. Mattiello, H. St\"ocker and W. Greiner:
Phys. Lett. B271(1991)37


\bibitem{correl} J.P. Sullivan, M. Berenguer, D.E. Fields, B.V. Jacak,
M. Sarabura, J. Simon-Gillo, H. Sorge, H. van Hecke, S. Pratt:
Phys. Rev. Lett. 70(1993)3000; Nucl. Phys. A566(1994)531c

 \bibitem{22} S.H. Kahana, Y. Pang and T.J. Schlagel:
 Nucl. Phys. A566(1993),465c and refs. therein


\bibitem{ART} B.A. Li,  C.M. Ko: Phys. Rev. C52 (1995) 2037



\bibitem{Keitz}
H. Sorge, A.v. Keitz, R. Mattiello, H. St\"ocker, W. Greiner:
 Phys. Lett. B 243(1990)7

\bibitem{21}
H.St\"ocker and W. Greiner: Phys. Rep. 137(1986)278;
H. Kruse, B.V. Jacak, and H. St\"ocker: Phys. Rev. Lett. 54(1985)289;
J.J. Molitoris, J.B. Hoffer, H. Kruse and H. St\"ocker:
Phys. Rev. Lett. 53(1984)899;
G. Buchwald, G. Graebner, J. Theis, J. Maruhn, and W. Greiner:
Phys. Rev. Lett. 52(1984)1594;
Ch. Hartnack, M. Berenguer, A. Jahns, A. v. Keitz,
R. Mattiello, A. Rosenhauer, J. Schaffner, T. Sch\"onfeldt, H. Sorge,
L. Winckelmann, H. St\"ocker, W. Greiner: Nucl. Phys. A538(1992)53c;


\bibitem{Daniel}
P. Danielewicz and Q. Pan: Phys. Rev. C46(1992)2002;
Q. Pan and P. Danielewicz: Phys. Rev. Lett. 70(1993)2062,3523

\bibitem{ogloblin} H. St\"ocker, A. Ogloblin, W. Greiner: Z. Phys.
A303(1981)259; S. Nagamiya, M.-C. Lemaire, E. Moeller, S. Schnetzer,
G. Shapiro, H. Steiner, and I. Tanihata: Phys. Rev. C24(1981)971


\bibitem{casflow} J.J. Molitoris, H St\"ocker, H.A. Gustafsson,
J. Cugnon, D. L'Hote: Phys. Rev. C33 (1986),867


\bibitem{experiment} K.-H. Kampert: J.Phys. G15(1989)691;
H.H. Gutbrod, K.H. Kampert, B.W. Kolb, A.M. Poskanzer, H.G. Ritter
and H.R. Schmidt: Phys. Lett. B216(1989)267


\bibitem{BMbodrum} 
J. Barrette for the E877 collaboration: Nucl. Phys. A590(1995)259c



\bibitem{asym} T. Abott for the E802 collaboration:
Phys. Rev. Lett. 70(1993)1393


\bibitem{di18} R. Mattiello, A. Jahns, H. Sorge, H. St\"ocker, W. Greiner:
Phys. Rev. Lett. 74(1995)2180


\bibitem{di117} A. Schwarzschild and C. Zupancic : Phys. Rev.
129(1963)854
\bibitem{di118}
H. Sato, K. Yazaki: Phys. Lett. B98(1981)153





\bibitem{di119} C.B. Dover, U. Heinz, E. Schnedermann, J. Zimanyi:
Phys. Rev. C44 (1991)1636


 \bibitem{coal1}
H. Gutbrodt, A. Sandoval, P. Johanssen, A. Poskanzer,
O. Gosset, W. Meyer, G. Westfall, R. Stock: Phys. Rev. Lett. 37 (1976)667


\bibitem{di134a}
H. Bando: Nuovo Cim. 102A (1989) 627;
H. Bando, M. Sano, J. Zofka and M. Wakai: Nucl. Phys. A501 (1989) 900

\bibitem{di134b}
F. Asai, H. Bando, M. Sano: Phys. Lett. 145B (1984) 19



\bibitem{di134c}
M. Wakai, H. Bando, M. Sano: Phys. Rev. C38(1988)748



\bibitem{di75} 
H. Kruse, B.V. Jacak, J.J. Molitoris, G.D. Westfall,
H. St\"ocker: Phys. Rev. C31(1985)1770 


  \bibitem{di131} A. Mekijan: Phys. Rev. Lett. 38(1977)640;
 Phys. Rev. C17(1978)1051


\bibitem{di132} P.R. Subramanian, L.P. Csernai, H. St\"ocker,
J.A. Maruhn, W. Greiner, H. Kruse: J. Phys. G: Nucl. Phys.
7(1981)L241

\bibitem{di133} D. Hahn und H. St\"ocker: Nucl. Phys. A476(1988)718


 \bibitem{di128} 
M. Gyulassy, K. Frankel and E.A. Remler: Nucl. Phys.
A402(1983)596



\bibitem{di129}
J. Aichelin, A. Rosenhauer, G. Peilert, H. St\"ocker, and W. Greiner:
Phys. Rev. Lett. 58(1987) 1926;
J. Aichelin, E.A. Remler: Phys. Rev. C35(1987) 1291


\bibitem{jamie} J.L. Nagle, S. Kumar, D. Kusnezov, H. Sorge, R. Mattiello:
Phys. Rev. C53(1996)367


\bibitem{bleicher} M. Bleicher, C. Spieles, A. Jahns, R. Mattiello,
H. Sorge, H. St\"ocker, W. Greiner: Phys. Lett. B361(1995)10


\bibitem{clscaling} H. Sorge, J.L. Nagle, B.S. Kumar:
Phys. Lett. B 355(1995)27


\bibitem{di127}
E.A. Remler and A.P. Sathe: Ann. of Phys. 91(1975)295
;E.A. Remler: Ann. of Phys. 95 (1975) 455
; E.A. Remler: Ann. of Phys. (NY) 136(1981)293


\bibitem{f9}
S. DeBenedetti: $Nuclear$ $Interactions$ (John Wiley 1964)



\bibitem{feedd} D.R. Tilley, H.R. Weller, H.H. Hasan: Nucl. Phys. 
A474(1987)1; D.R. Tilley, H.R. Weller, G.M. Hale: Nucl. Phys. A541(1992)1


\bibitem{f20} 
R.G. Sachs and M. Goeppert-Mayer: Phys. Rev. 53(1938)991


\bibitem{f22} L. Hulthen: Ark. Mat. Ast. Fys. 28, No. 5



\bibitem{di149}
P.B. Siegel and M. Farrow-Reid: Am. J. Phys. 58(1990) 1016


\bibitem{di140} G.R. Shin and J. Rafelski: 
J. Phys. G: Nucl. Part. Phys. 16(1990)L187



\bibitem{di144}
H.J. Mang: Phys. Rev. 119 (1960) 1069



\bibitem{di145} R. Hofst\"atter: Rev. Mod. Phys. 28(1956)214



\bibitem{di70} 
J. Carlson: Phys. Rev. C38 (1988) 1879



\bibitem{di71} 
C.R. Chen, G.L. Payne, J.L. Friar, B.F. Gibson: 
Phys. Rev. C33 (1986)1740




 \bibitem{4}
 B.D. Serot and J.D. Walecka: Adv. Nucl. Phys. 15 (1986);
 J. Theis et al.: Phys. Rev. D28 (1983)2286


\bibitem{pa1} E.D. Cooper, B.C. Clark, R. Kozack, S. Shim, S. Hama,
J.I. Johansson, H.S. Sherif, R.L. Mercer, B.D. Serot:
Phys. Rev. C36(1987)2170


\bibitem{pa2} M. Jaminon, C. Mahaux, P. Rochus:
Nucl. Phys. A365(1981)371

\bibitem{pa3} B. Ter Haar and R. Malfliet: Phys. Lett. B172(1986)10

\bibitem{pa4} T.L. Ainsworth, E. Baron, G.E. Brown, J. Cooperstein,
M. Prakash: Nucl. Phys. A464(1987)740



\bibitem{25} F. de Jong and R. Malfliet: Phys. Rev.
C46(1992)2567


\bibitem{QGP} J. Ellis, J. Kapusta, K. Olive: Phys. Lett. B273(1991)122


 \bibitem{10}
 H. Sorge, R. Mattiello, H. St\"ocker and W. Greiner:
Phys. Rev. Lett. 68(1992)286.


\bibitem{ropes} H. Sorge: Phys. Rev. C52(1995)3291


\bibitem{HSNEW} H. Sorge: manuscript in preparation


\bibitem{di125} J.L. Nagle, B.S. Kumar, M.J. Bennett, G.E. Diebold,
J.K. Pope, H. Sorge, J.P. Sullivan:
Phys. Rev. Lett. 73 (1994) 1219




\bibitem{f12} 
T. Abbott for the E802 collaboration: Phys. Rev. C50 (1994) 1024



\bibitem{AGSclust} D. Beavis for the E878 collaboration:
Phys. Rev. Lett. 75(1995)3078

\bibitem{gillo} J. Gillo et al.: Nucl. Phys. A590(1995),483c

\bibitem{bejin} Ziping Chen for the E802 collaboration: Contrib. to
the First Int. Conf. on Frontiers of Physics, Shantou, China, August 1995



\bibitem{di52} B. Moscowitz, M. Gonin, F. Videbaek, H. Sorge, R. Mattiello:
Phys. Rev. C 51(1995)310


\bibitem{siemens} P.J. Siemens and J.O. Rasmussen: Phys. Rev. Lett.
42(1978)880

\bibitem{eostrans} M.A. Lisa for the EOS collaboration: Phys. Rev. Lett.
75(1995) 2662

\bibitem{konopka} J. Konopka: PhD-thesis, J.W. Goethe Universit\"at
Frankfurt am Main, 1995

 \bibitem{15}
 K.S. Lee and U. Heinz: Z. Phys. C48(1990)525;
K.S. Lee, U. Heinz and E. Schnedermann: 
Z. Phys. C48 (1990) 525


\bibitem{di59} E. Schnedermann and U. Heinz: Phys. Rev. Lett.
69 (1992) 2908


\bibitem{stachel} P. Braun-Munzinger, J. Stachel, J.P. Wessels, N. Xu:
Phys. Lett. B365(1996)1

\bibitem{greek} A.D. Panagiotou, G. Mavromanolakis, J. Tzoulis:
Phys. Rev. C53(1996)1353



\bibitem{di409}
M. Hofmann, R. Mattiello, H. Sorge, H. St\"ocker and W. Greiner:
Phys. Rev. C51(1995) 2095
 

\bibitem{BSW} G.E. Brown, J. Stachel, G.M. Welke: 
Phys. Lett. B253(1991)19


\bibitem{Gutbrod} H.A. Gustafsson, H.H. Gutbrod, J. Harris,
B.V. Jacak, K.H. Kampert, B.Kolb, A.M. Poskanzer, H.G. Ritter and
H.R. Schmidt: Mod. Phys. Lett. A3(1988)1323

\bibitem{Strott} N.S. Amelin, E.F. Staubo, L.P. Csernai, V.D. Toneev,
K.K. Gudima, D. Strottman: Phys. Rev. Lett. 67(1991)1523



\bibitem{eosflow} M. D. Partlan for the EOS collaboration:
Phys. Rev. Lett. 75(1995)2100


\bibitem{di402} A. Jahns, Chr. Spieles, H. Sorge, 
H. St\"ocker and W. Greiner: Phys. Rev. Lett. 72(1994) 3464



\bibitem{di198}
S.A. Bass, R. Mattiello, H. St\"ocker, W. Greiner and Ch. Hartnack:
Phys. Lett. B302 (1993) 381


\bibitem{densflow} S.A. Bass, C. Hartnack, H. St\"ocker, W.Greiner:
Phys. Rev. C51(1995)3343


\bibitem{Li} B.A. Li, C.M. Ko: Phys. Rev. C53(1996)R22


\bibitem{wessels} J. Wessels (E877-collab.): Proc 11$^{th}$ Winter
Workshop on Nuclear Dynamics, Key West, in press


\end{thebibliography}
\end{document}